\documentclass[final,5p,times,twocolumn]{elsarticle}

\usepackage{amssymb}
\usepackage{color}
\usepackage{amsmath}
\usepackage{multirow}
\usepackage{booktabs}
\usepackage{bm}

\usepackage[pdfstartview=FitH, colorlinks,
 linkcolor={red!60!black!},
 anchorcolor={green!80!black!},
 urlcolor={blue!80!black!},
 citecolor={blue!50!black!}]{hyperref}
\usepackage[table]{xcolor}

\journal{Physics Letters B}

\begin{document}


\begin{frontmatter}



\title{Continuum and three-nucleon force in Borromean system: The $^{17}$Ne case}


\author[ad1]{Y. Z. Ma}
\author[ad1]{F. R. Xu\corref{cor1}}
\author[ad2,ad3]{N. Michel}
\author[ad1]{S. Zhang}
\author[ad1]{J. G. Li}
\author[ad1]{B. S. Hu}
\author[ad4]{L. Coraggio}
\author[ad4,ad5]{N. Itaco}
\author[ad4]{A. Gargano}

\address[ad1]{School of Physics, and State Key Laboratory of Nuclear Physics and Technology, Peking University, Beijing 100871, China}
\address[ad2]{Institute of Modern Physics, Chinese Academy of Sciences, Lanzhou 730000, China}
\address[ad3]{School of Nuclear Science and Technology, University of Chinese Academy of Sciences, Beijing 100049, China}
\address[ad4]{Istituto Nazionale di Fisica Nucleare, Complesso Universitario di Monte S. Angelo, Via Cintia, I-80126 Napoli, Italy}
\address[ad5]{Dipartimento di Matematica e Fisica, Universita` degli Studi della Campania “Luigi Vanvitelli”, viale Abramo Lincoln 5 - I-81100 Caserta, Italy}
\cortext[cor1]{frxu@pku.edu.cn}

\begin{abstract}
Starting from chiral two-nucleon (2NF) and chiral three-nucleon (3NF) potentials, we present a detailed study of $^{17}$Ne, a Borromean system, with the Gamow shell model which can capture continuum effects.
More precisely, we take advantage of the normal-ordering approach to include the 3NF and the Berggren representation to treat bound, resonant and continuum states on equal footing in a complex-momentum plane.
In our framework, 3NF is essential to reproduce the Borromean structure of $^{17}$Ne, while the continuum is more crucial for the halo property of the nucleus.
The two-proton halo structure is demonstrated by calculating the valence proton density and correlation density.
The astrophysically interesting $3/2^-$ excited state has its energy above the threshold of the proton emission, and therefore the two-proton decay should be expected from the state.
\end{abstract}


\end{frontmatter}



\section{Introduction}
One of the main challenges in modern nuclear theory is to understand nuclei close to proton and neutron driplines,
which is also the focal point of current and next-generation rare-isotope-beam (RIB) facilities in experimental programs.
Due to extreme proton-to-neutron imbalance, the nuclei exhibit exotic phenomena such as halo and Borromean structures, making them become ideal laboratories to study modern nucleon interactions, e.g., nuclear forces from chiral effective field  theory \cite{Weinberg:1990rz,RevModPhys.81.1773, Machleidt2002} and to test the reliability of existing many-body methods.
One may think of the neutron halos in $^{6,8}$He \cite{Wang2004, Mueller2007}, $^{11}$Be \cite{Kelley1995} and $^{11}$Li \cite{Tanihata1985}.
The neutron dripline has been studied from the experimental point of view in the $A \simeq 20$ region, as it has been reached in the chains of fluorine and neon isotopes \cite{F-Ne-2019}.
One is therein at the limit of experimental possibilities, as the neutron dripline is out of reach for heavier isotopes.
Conversely, because of the Coulomb repulsion, the proton dripline is not so far from the stability line compared with the neutron dripline, and has been reached experimentally up to Z$\approx$90 more than a decade ago \cite{Thoennessen_2004}.
However, when focusing on light proton-rich nuclei where the Coulomb barrier is not so high and the continuum effect is prominent, interesting phenomena (like halo and Borromean structures) may emerge, similarly to the neutron-rich side of the nuclear landscape.
Among these nuclei, $^{17}$Ne is particular and attracts a lot of interest, both theoretically \cite{GARRIDO200485, Casal2016, Parfenova2018} and experimentally \cite{Tanaka2010,Sharov2017,Charity2018}.

The $^{17}$Ne nucleus is located on the proton dripline and is a Borromean nucleus with an unbound subsystem $^{16}$F.
$^{17}$Ne may bear a similarity to the two-neutron halo nucleus $^{6}$He which can be described in a cluster picture of the three-body $^{4}$He + 2n system.
The question about two-proton halo in $^{17}$Ne was formulated in Ref \cite{Zhukov1995} but still not confirmed \cite{GARRIDO200485, Tanaka2010, Casal2016}.
This problem is enhanced by the prediction that $^{17}$Ne might be the only candidate to unveil the two-proton halo structure, as the Coulomb interaction should impose stronger confinement in heavier proton-rich systems \cite{Sharov2017}.
Moreover, the two-proton decay from $^{17}$Ne low-lying states to $^{15}$O (which is a ``waiting point" in the astrophysical rp-process) has a strong connection to the hot Carbon-Nitrogen-Oxigen (CNO) cycle \cite{CNO-1, CNO-2, Casal2016, Sharov2017},
so that the theoretical study of $^{17}$Ne is also interesting for astrophysics.
Within this context, the radioactive properties of $^{17}$Ne have been studied,
using a three-body model \cite{Casal2016, Parfenova2018} and in the shell model embedded in the continuum \cite{Charity2018}.

In theoretical studies of weakly-bound and unbound nuclei, due to the low particle-emission thresholds, it is crucial to consider the strong coupling to resonance and continuum.
Coupling to the continuum is indeed responsible of the bound character of Borromean nuclei, for example.
An elegant treatment of resonance and continuum effects is based on the Berggren basis \cite{Berggren1968}.
The Berggren basis generalizes one-body Schr\"odinger equation to a complex-$k$ plane and generates bound, resonant and non-resonant continuum single-particle (SP) states naturally.
By representing the nuclear Hamiltonian in the Berggren basis, one can conveniently calculate the many-body states of weakly bound and unbound nuclei, as they become the eigenstates of a complex symmetric matrix in the Berggren formalism.
This is the object of the Gamow shell model (GSM) which has been developed to study the nuclear structure of weakly bound and resonance nuclei.
Phenomenological interactions have been used along with GSM \cite{IdBetan2002, Michel2002, Michel2003, Michel2019,Papadimitriou2011, Michel2020}.
More realistic approaches have been introduced in GSM, e.g., the no-core GSM \cite{Papadimitriou2013}, the core GSM with realistic nuclear forces \cite{Hagen2006, Tsukiyama2009, Sun2017, Hu2020, zhuo2020}.
The Berggren basis has also been applied in the coupled cluster (CC), named the complex CC framework, to study helium \cite{HAGEN2007169} and neutron-rich oxygen \cite{Hagen2012} isotopes, as well as the positive-parity states of $^{17}$F \cite{Hagen2010}.
In Ref. \cite{Hu2019}, the Berggren basis was applied in the in-medium similarity renormalization group (Gamow IM-SRG).

Meanwhile, many theoretical works \cite{Navratil2007, Otsuka2010,Holt2013, PhysRevLett.113.142501, PhysRevLett.113.142502} have shown the importance of taking into account the three-nucleon force (3NF) in nuclear structure calculations with realistic potentials.
It is worth pointing out that the chiral perturbation theory (ChPT) generates
nuclear two-, three- and many-body forces on an equal footing, since most interaction vertices appear in 3NF also occur in 2NF.
One significant effect from 3NF is that it can provide a strong repulsive contribution \cite{Otsuka2010} which can resolve the long-standing over-binding problem in heavy nuclei with most realistic two-body forces.
3NF has been recently introduced in GSM and applied to to the study of neutron-rich nuclei \cite{zhuo2020}.
We have recognized the necessity of combining continuum and 3NF together for understanding nuclei around drip line, in order to get reliable results and predictions for weakly-bound Borromean systems.

The aim of this work is then to study the $^{17}$Ne isotope in GSM, for which both chiral two-nucleon force (2NF) and 3NF will be used in the Hamiltonian.
2NF is considered at next-to-next-to-next-to-leading order (N$^3$LO), whereas 3NF is calculated at next-to-next-to-leading order (N$^2$LO).
We can then properly investigate the combined roles of 2NF, 3NF and continuum degrees of freedom in the halo formation occurring in $^{17}$Ne.

\section{Outline of calculations}
We start from the chiral N$^3$LO potential derived by Entem and Machleidt \cite{Machleidt2002} as the 2NF and a chiral N$^2$LO 3NF as the 3NF.
The chiral N$^2$LO 3NF consists of three components, namely the two-pion (2$\pi$) exchange $V^{(2\pi)}_{\rm 3N}$, the one-pion (1$\pi$) exchange plus contact $V^{(1\pi)}_{\rm 3N}$ and the contact term $V^{(\rm ct)}_{\rm 3N}$.
It should be pointed out that the low-energy constants (LECs) $c_1$, $c_3$ and $c_4$ appearing in $V^{(2\pi)}_{\rm 3N}$ are the same as those in 2NF, so their values are already fixed during the construction of the N$^3$LO two-nucleon potential.
However, there are still two LECs $c_D$ and $c_E$ characterizing one-pion exchange and contact term, which cannot be constrained by two-body observables and need to be determined by reproducing observables in systems with mass number $A > 2$.
We adopt the same values of $c_D=-1$ and $c_E=-0.34$ given in Refs. \cite{zhuo2018, zhuo2019, zhuo2020}.

For the GSM calculation, similar to realistic shell model (RSM) \cite{Coraggio2009-1,Coraggio2009-2}, an auxiliary one-body potential $U$ is introduced into Hamiltonian, which decomposes the intrinsic Hamiltonian of an A-nucleon system into a one-body term $H_0$ and a residual interaction $H_1$, as follows,
\begin{equation}\label{eq:hm1}
 \begin{aligned}
   H&=\sum_{i<j}\frac{(\boldsymbol{p}_i-\boldsymbol{p}_j)^2}{2mA}+\hat{V}_{\text{NN}}+\hat{V}_{\text{3N}}\\
    &=\sum_{i=1}^A\left(\frac{p_i^2}{2m}+U\right)
      +\sum^A_{i<j}\left({V}_{\text{NN}}^{(ij)}-U-\frac{p_i^2}{2mA}-\frac{\boldsymbol{p_i}\cdot \boldsymbol{p_j}}{mA}\right)\\
    &+\sum^A_{i<j<k}{V}_{\text{3N}}^{(ijk)}\\
    &=H_0+H_1,
 \end{aligned}
\end{equation}
with $H_0=\sum_{i=1}^A(\frac{p_i^2}{2m}+U)$ having a one-body form and describing the independent motion of the nucleons.
In the present calculations, $U$ is taken as the Woods-Saxon (WS) potential of the $^{14}$O core.
Due to the explosive dimension of the shell model with full inclusion of 3NF, particularly when continuum states are included, we employed the normal-ordering approximation \cite{Roth2012, zhuo2018, zhuo2020} to introduce the contribution of 3NF into our calculations.
It has been shown that the 3NF normal-ordering approximation with neglecting the residual three-body term works well in nuclear structure calculations \cite{Roth2012}.
The normal-ordered two-body term can be written as
\begin{equation}\label{eq:hm2}
 \begin{aligned}
   \hat{V}_{\rm 3N}^{\rm (2B)}
   &=\frac{1}{4}\sum_{\substack{ijkl}}\langle{ij}\vert{V_{\rm 3N}^{\rm (2B)}}\vert{kl}\rangle
      \{{a}^{\dagger}_ia^{\dagger}_ja_la_k\}\\
   &=\frac{1}{4}\sum_{\substack{ijkl}}\sum_{h\in{\rm core}}\langle{ijh}\vert{V_ {\rm 3N}}\vert{klh}\rangle\{{a}^{\dagger}_ia^{\dagger}_ja_la_k\},
  \end{aligned}
\end{equation}
where $\langle{ijh}\vert{V_{\rm 3N}}\vert{klh}\rangle$ and $\langle{ij}\vert{V^{\rm (2B)}_{\rm 3N}}\vert{kl}\rangle$ are the antisymmetrized matrix elements of the 3NF and the normal-ordered two-body term of the 3NF, respectively. $a_i^\dagger$ ($a_i$) stands for the particle creation (annihilation) operator with respect to the nontrivial vacuum.
The symbol \{...\} means that the creation and annihilation operators in brackets are normal-ordered.
For the calculation of $^{17}$Ne, the closed-shell nucleus $^{14}$O is chosen as the core, with its ground-state Slater determinant as the reference state for the normal-ordering decomposition.
We firstly calculate antisymmetrized N$^2$LO 3NF matrix elements in the Jacobi-HO (harmonic oscillator) basis in the momentum space, and then the normal-ordered 3NF two-body matrix elements are added to the N$^3$LO 2NF matrix elements.
The full matrix elements are transformed into the Berggren basis by computing overlaps between the HO and Berggren basis wave functions.
More details can be found in Ref. \cite{zhuo2020}.
The maximum shell number $N_{shell}=2n+l+1=23$ with a limit of $l\le 4$ ($n$ and $l$ standing for the HO node and orbital angular momentum, respectively) is taken as the truncation in the HO basis, and 40 discretized scattering states are used in Berggren continuum contours.
Convergence has been tested in Ref.\cite{Sun2017}.
In principle, one should also consider the effect of center-of-mass (CoM) motion because the wave functions are written in the laboratory coordinates, though Hamiltonian (\ref{eq:hm1}) is purely intrinsic.
However, it has been observed that the CoM effect can be neglected for low-lying states \cite{Sun2017, Hu2019}.

Since we choose the closed-shell $^{14}$O as core, the proton and neutron SPs in the model space actually correspond to the energy spectra of $^{15}$F and $^{15}$O.
The well bound $0p_{1/2}, 0d_{5/2}, 1s_{1/2}$ and $0d_{3/2}$ orbitals are selected to be the active neutron model space.
Due to the unbound character of the ground state of $^{15}$F, whose ground $1/2^+$ and first $5/2^+$ excited states are respectively depicted by the $1s_{1/2}$ and $0d_{5/2}$ proton orbitals,
it is necessary to include the $1s_{1/2}$ and $0d_{5/2}$ orbitals, and associated scattering states of the same partial waves in the proton model space.
As the SP excited $3/2^+$ state of $^{15}$F is high in energy, it is not necessary to include the proton $d_{3/2}$ partial wave in the model space.

Usually the Berggren basis is generated by the WS potential including a spin-orbit coupling.
In this work, the WS parameters are based on the `universal' one given in Ref. \cite{Dudek1981}.
However, to obtain the inverse positions of the proton $1s_{1/2}$ and $0d_{5/2}$ orbitals in $^{15}$F, we reduce the spin-orbital coupling strength to 2 MeV and reduce the depth parameter by 2 MeV for the proton WS potential.
In physics, the need to reduce the spin-orbit coupling strength of the WS potential is due to the fact that 3NF contributes largely to the spin-orbit component of the residual interaction (see comments in Refs. \cite{zhuo2018, zhuo2019}).
Then, we obtain proton single-particle energies: $\tilde{e}=0.456-0.002i$ MeV for $1s_{1/2}$, and $\tilde{e}=1.957-0.034i$ MeV for $0d_{5/2}$ ($\tilde{e}_n=e_n-i\gamma_n/2$, and $\gamma_n$ standing for the resonance width), which are close to experimental single-particle levels in $^{15}$F.
Therefore, the choice of the model space is the neutron bound $\nu\{0p_{1/2}, 0d_{5/2}, 1s_{1/2}, 0d_{3/2}\}$ and the proton resonances $\pi\{1s_{1/2}, 0d_{5/2}\}$ plus continua $\pi\{s_{1/2}, d_{5/2}\}$.
To speed up the convergences of many-body calculations, we soften the 2NF by using $V_{\text{low-}k}$ method \cite{Bogner2003}, with a cutoff $\Lambda=2.6$ fm$^{-1}$ same as in Ref. \cite{Sun2017}.
Next, we construct the effective interaction in the model space, using the many-body perturbation theory.
More precisely, we exploit the $\hat Q$-box folded diagrams \cite{Kuo197165} to the complex-$k$ space with the extended Kuo-Krenciglowa (EKK) method \cite{Takayanagi201161},
during which contributions from the core polarization and other continuum partial waves are taken into account \cite{Sun2017,zhuo2020}.
Due to the explosion of the model dimension when continuum included, the perturbative expansion of the $Q$-box is at the second-order level (computationally prohibitive to go to third order).
At last, the complex symmetric non-Hermitian GSM Hamiltonian is diagonalized in the GSM space given above, by using the Jacobi-Davidson method in the m-scheme.

\section{Results}
\begin{figure}[!ht]
\begin{center}
\includegraphics[width=0.45\textwidth]{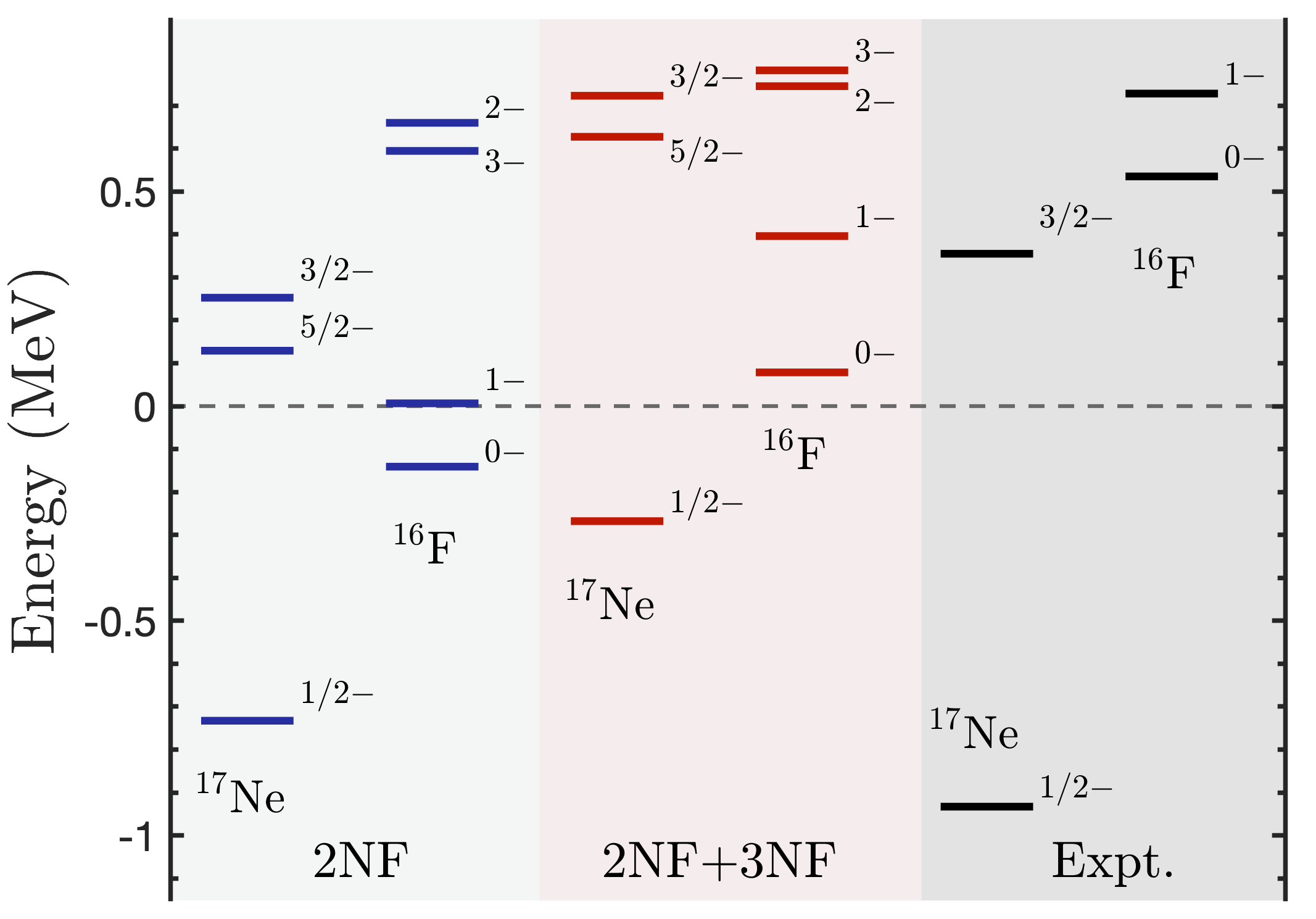}
\caption{\label{fig:Spec}Calculated spectra of $^{17}$Ne along with its isotone $^{16}$F, with respect to $^{15}$O. Blue and red lines are the GSM calculations with only 2NF interaction and 2NF+3NF interaction respectively.
Experimental data are taken from Refs. \cite{Audi2012, Sharov2017}.}
\end{center}
\end{figure}
The calculated low-lying spectra of $^{17}$Ne and its isotone $^{16}$F are presented in Fig. \ref{fig:Spec}, with respect to the $^{15}$O ground-state energy.
The $^{15}$O ground-state energy actually is the ${\nu}0p_{1/2}$ position with respect to the $^{14}$O core.
In our framework, the valence proton SP states are all unbound, which means that only the naive occupation of protons in model space can not give bound $^{17}$Ne and its bound nature should come from the correlation of valence particles.
The comparison of the blue and red lines in Fig. \ref{fig:Spec} shows that 3NF lifts the whole spectra of $^{17}$Ne and $^{16}$F, making $^{16}$F unbound and $^{17}$Ne with a Borromean character.
This Borromean structure can be seen more clearly in Table \ref{tab:table1}, where the calculations with 3NF provide an unbound $^{16}$F and a bound $^{17}$Ne.
\begin{table}[h]\small
\center
\setlength{\belowcaptionskip}{5pt}%
\caption{\label{tab:table1} Calculated ground-state energies of $^{17}$Ne and $^{16}$F (with respect to $^{15}$O) with and without 3NF. The experimental data are taken from Ref. \cite{Audi2012}.}
\renewcommand\arraystretch{1.2}
\setlength{\tabcolsep}{4.2mm}
\begin{tabular}{llll}
\hline
\hline
 E(MeV)       & 2NF only        & 2NF+3NF   & Expt. \\ \hline
$^{17}$Ne    & $-0.73$   & $-0.26$ & $-0.95$ \\
$^{16}$F     & $-0.14$   & $+0.08$ & $+0.54$ \\
\hline
\hline
\end{tabular}
\end{table}
In our previous work \cite{zhuo2020}, we have found that all of the three components, 2$\pi$-exchange $V^{(2\pi)}_{\rm 3N}$, 1$\pi$-exchange $V^{(1\pi)}_{\rm 3N}$ and contact $V^{(\text{ct})}_{\rm 3N}$,  have significant contributions but they behave differently.
The repulsive contact term $V^{(\text{ct})}_{\rm 3N}$ has a similar absolute value as the attractive 1$\pi$-exchange $V^{(1\pi)}_{\rm 3N}$ and their net effect is almost cancelled out, leaving the long-range repulsive two-pion exchange $V^{(2\pi)}_{\rm 3N}$ dominant \cite{zhuo2020}.
Moreover, 3NF has a more remarkable repulsive effects in $^{17}$Ne than in $^{16}$F, which makes a smaller gap of ground-state energies between the two isotones.
As well known, the 3NF contribution increases with increasing the number of valence particles \cite{zhuo2020}.
The differences seen in Fig. \ref{fig:Spec} and Table \ref{tab:table1} between calculations and data may be due to the lack of correlations coming from higher-order contributions of the Q-box perturbative expansion.
In a strict single-particle formulation of the shell model, there should not exist explicitly three-nucleon effects in $^{16}$F with only one neutron and one proton in the model space.
However, when we construct the effective interaction in the model space, the ``core" is not frozen and the correlation from nucleons in the core space can be ``folded" into two-body interaction by the folded-diagram procedure.
That means that the 3NF effect in $^{16}$F can be traced back to the nucleons in the core.
Besides, we found strong configuration mixing in the ground state $^{17}$Ne, with a $54\%$ $s$-wave component of $\pi 1s_{1/2}^2\otimes\nu 0p_{1/2}^1$.
This is consistent with the result discussed in Ref. \cite{Parfenova2018}.

\begin{figure}[!ht]
\begin{center}
\includegraphics[width=0.48\textwidth]{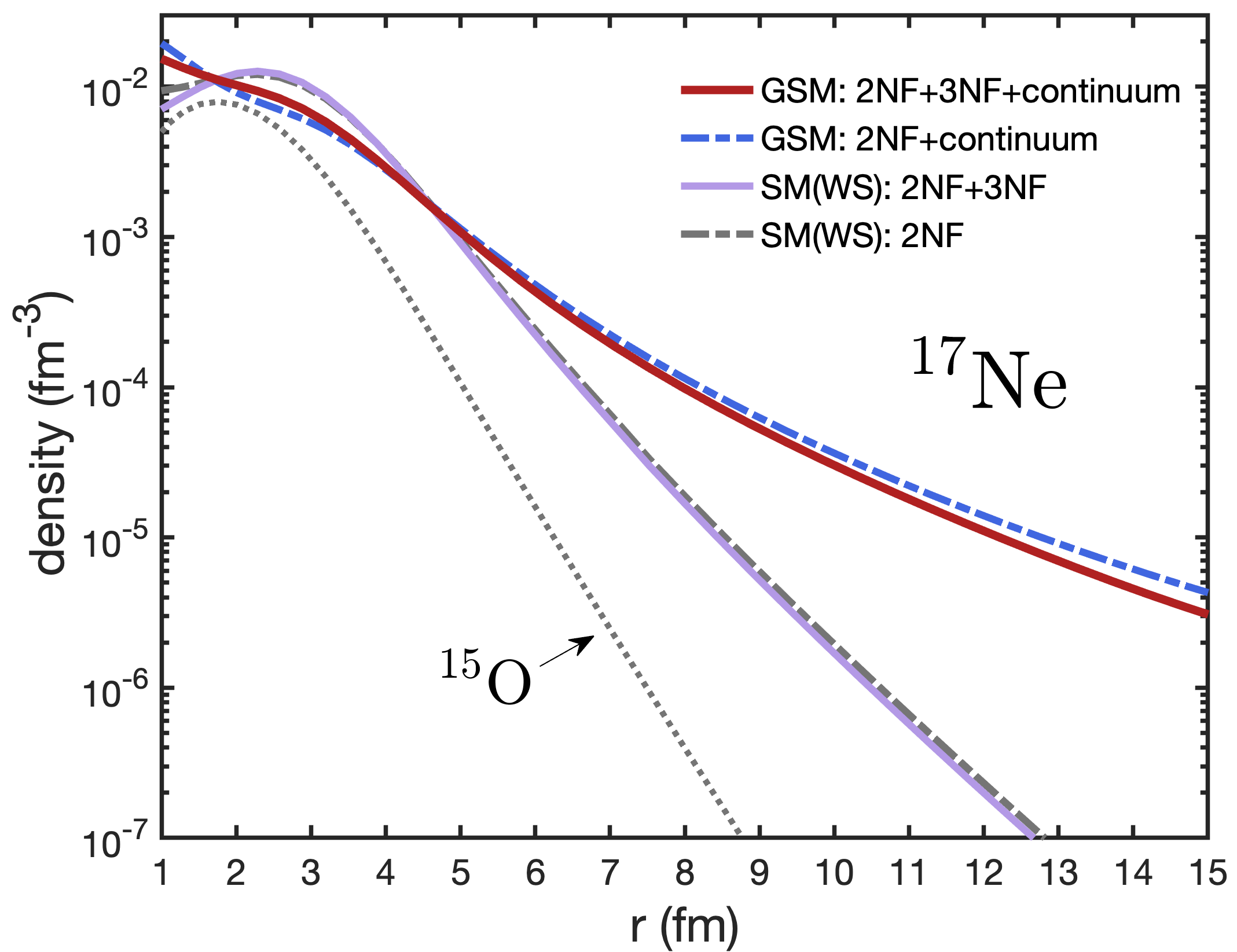}
\caption{\label{fig:density} The $^{17}$Ne density in the valence space, calculated by GSM with 2NF only (blue dot-dashed line) and 2NF+3NF (red line).
SM(WS) stands for the shell-model calculations without continuum performed in the non-continuum discrete WS basis, with 2NF+3NF (purple line) and 2NF only (grey dot-dashed line).
The $^{15}$O density (black dash line) in valence space is also displayed for comparison.}
\end{center}
\end{figure}
The small two-proton separation energy and the large weight of the proton $s$-wave configuration in the ground state is a typical signature of the presence of a halo in $^{17}$Ne.
In order to corroborate this assumption, we calculated the one-body density of $^{17}$Ne by the GSM (shown in Fig. \ref{fig:density}).
By comparing the density of $^{17}$Ne with that of $^{15}$O, we find that the $^{17}$Ne density has a long ``tail'' which is a direct evidence to support the halo nature of $^{17}$Ne.
The comparison between calculations with and without 3NF shows that 3NF does not have significant influence on the density distribution.
Furthermore, to underline the continuum effect within the present model, we have performed the shell-model calculation without the inclusions of the continuum.
In this calculation, the WS potential is solved in the HO basis (instead of the Berggren representation), which gives non-continuum discrete WS single-particle states, see \cite{zhuo2020} for more details.
We see from Fig. \ref{fig:density} that the calculated $^{17}$Ne density (with or without 3NF) decreases rapidly at large distance when excluding the continuum.
This can be explained in two aspects: 1) the $\pi 1s_{1/2}^2\otimes\nu 0p_{1/2}^1$ configuration
is reduced to about 20\% when excluding the continuum; 2) the $s$-wave expanded by the HO basis can not carry enough density information at large distance due to the spatially-localized property of the HO basis.
Therefore, the continuum effect is crucial to describe the halo structure of $^{17}$Ne.

\begin{figure}[!ht]
\begin{center}
\includegraphics[width=0.48\textwidth]{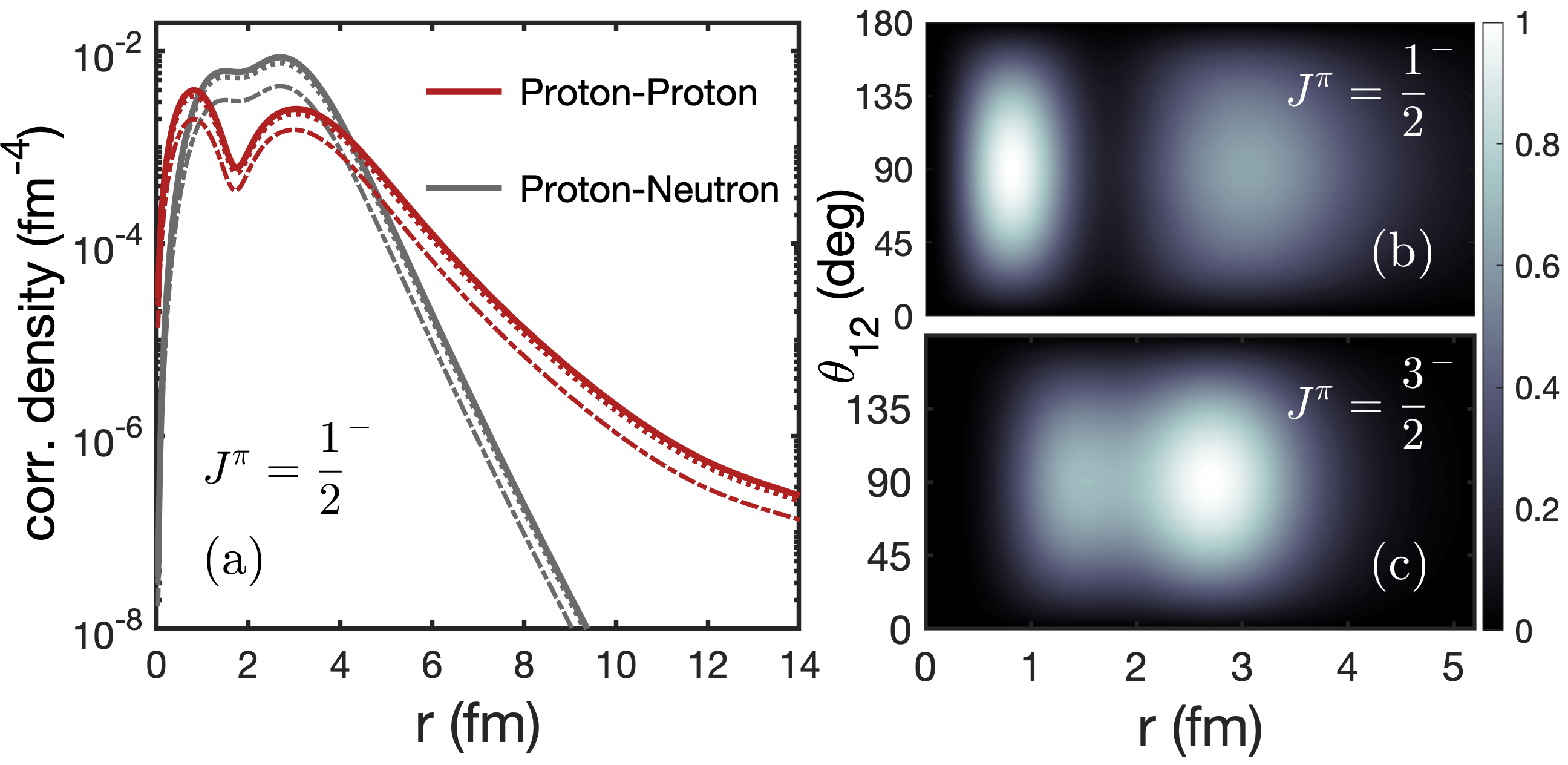}
\caption{\label{fig:corr_density} Calculated proton-proton and proton-neutron correlation densities of the $^{17}$Ne ground state (a) with $\theta_{12}$ equal to $ 30^{\circ}$, $90^{\circ}$, $120^{\circ}$ by dot-dashed, solid and dot lines, respectively.
The two-dimension proton-proton correlation density $\rho (r, \theta_{12})/\rho_0$ (normalized with the maximum value $\rho_0$ of the correlation density) is also plotted for the $1/2^-$ ground state (b) and $3/2^-$ excited state (c). }
\end{center}
\end{figure}
In order to see the inner structure of the halo, we calculate the correlation densities of $^{17}$Ne and illustrate in Fig. \ref{fig:corr_density}.
The definition of the correlation density is defined as \cite{Papadimitriou2011, Michel2020},
\begin{equation}
\rho\left(r, \theta_{12}\right)
=\langle\Psi\vert{\frac{1}{r^2}{\delta\left(r-r_{1}^{\prime}\right)}{\delta\left(r-r_{2}^{\prime}\right)}\delta\left(\theta_{12}-\theta_{12}^{\prime}\right)}\vert{\Psi}\rangle,
\end{equation}
where $r_1^\prime$ ($r_2^\prime$) is the radial coordinate of the first (second) nucleon, $\theta^\prime_{12}$ represents the angle between the two nucleons relative to the origin of coordinates, and $\vert{\Psi}\rangle$ is the wave function of the state.
The Jacobian induced by angular dependence is implicitly included in $\rho(r,\theta_{12})$, see Ref. \cite{Michel2020}.
In panel (a) of Fig. \ref{fig:corr_density}, we show the proton-proton and proton-neutron correlation densites $\rho(r,\theta_{12})$ in three different $\theta_{12}$ angles ($ 30^{\circ}$, $90^{\circ}$ and $120^{\circ}$).
It is clearly seen that the proton-proton correlation density has a ``long tail", which supports the two-proton halo structure of the $^{17}$Ne ground state.
From panel (b) of Fig. \ref{fig:corr_density}, one can see that the two-proton correlation density of the $1/2^-$ ground state concentrates mainly in an area close to the core at $r\sim 1$ fm with an angle $\theta_{12} \sim 90^\circ$ between the two protons, while the rest of the distribution is around $r\sim 2$-$4$ fm with $\theta_{12}\sim 90^\circ$.
The two-proton decay of the excited $J^{\pi}=3/2^-$ state is an important issue for solving the bypass problem of the $^{15}$O waiting point in nuclear astrophysics \cite{Sharov2017}, since the radioactive absorption of two protons is known to be a possible bypath for this waiting point \cite{CNO-1}.
Then we also demonstrate the two-proton correlation density of the $J^{\pi}=3/2^-$ state in the panel (c) of Fig. \ref{fig:corr_density}.
The two-proton correlation density of the $3/2^-$ excited state concentrates mainly around $r\sim 3$ fm with an angle $\theta_{12} \sim 90^\circ$, meaning that the two protons in the excited state stay in a farther area from the core than in the ground state.
The $3/2^-$ excited state is above the threshold of the two-proton decay.
We may expect a two-proton decay from the $^{17}$Ne $3/2^-$ excited state.

\section{Conclusions}
In conclusion, starting from a chiral 2NF and a chiral 3NF, we have presented the results of the GSM calculations for $^{17}$Ne and investigated the roles played by the continuum and 3NF in several aspects.
More specifically, we take the advantage of the normal-ordering approach to include 3NF effects and the Berggren representation to treat bound, resonant and continuum states on equal footing in a complex-$k$ plane.
From the calculated results, we find that both continuum and 3NF are crucial to describe the Borromean halo nucleus, $^{17}$Ne.
In particular, 3NF is essential for binding energy and the continuum is more crucial for density.
The repulsive 3NF raises the energy of $^{16}$F over the threshold of the proton emission, which leads to a Borromean structure of $^{17}$Ne.
The nucleus is further investigated by calculating one-body density and two-nucleon correlation density, showing a two-proton halo structure.
The $3/2^-$ excited state lies above the threshold of the proton emission, therefore a two-proton decay should be possible, which is interesting also in astrophysics.

\section{Acknowledgements}
This work has been supported by the National Key R\&D Program of China under Grant No. 2018YFA0404401; the National Natural Science Foundation of China under Grants No. 11835001, No. 11921006, No. 11975282 and No. 11435014; China Postdoctoral Science Foundation under Grant No. BX20200136; the State Key Laboratory of Nuclear Physics and Technology, Peking University under Grant No. NPT2020ZZ01; the Strategic Priority Research Program of Chinese Academy of Sciences, Grant No. XDB34000000; and the CUSTIPEN (China- U.S. Theory Institute for Physics with Exotic Nuclei) funded by the U.S. Department of Energy, Office of Science under Grant No. de-sc0009971. We acknowledge the High-Performance Computing Platform of Peking University for providing computational resources.


\section*{References}

\bibliographystyle{elsarticle-num_noURL}
\bibliography{references-plb}

\begin{thebibliography}{10}
\expandafter\ifx\csname url\endcsname\relax
  \def\url#1{\texttt{#1}}\fi
\expandafter\ifx\csname urlprefix\endcsname\relax\def\urlprefix{URL }\fi
\expandafter\ifx\csname href\endcsname\relax
  \def\href#1#2{#2} \def\path#1{#1}\fi

\bibitem{Weinberg:1990rz}
S.~Weinberg, Phys. Lett. B251 (1990) 288--292.

\bibitem{RevModPhys.81.1773}
E.~Epelbaum, H.-W. Hammer, U.-G. Mei\ss{}ner,
  \href{https://link.aps.org/doi/10.1103/RevModPhys.81.1773}{Rev. Mod. Phys.}
  81 (2009) 1773--1825.

\bibitem{Machleidt2002}
D.~R. Entem, R.~Machleidt,
  \href{https://link.aps.org/doi/10.1103/PhysRevC.66.014002}{Phys. Rev. C} 66
  (2002) 014002.

\bibitem{Wang2004}
L.-B. Wang, P.~Mueller, K.~Bailey, G.~W.~F. Drake, J.~P. Greene, D.~Henderson,
  R.~J. Holt, R.~V.~F. Janssens, C.~L. Jiang, Z.-T. Lu, T.~P. O'Connor, R.~C.
  Pardo, K.~E. Rehm, J.~P. Schiffer, X.~D. Tang,
  \href{https://link.aps.org/doi/10.1103/PhysRevLett.93.142501}{Phys. Rev.
  Lett.} 93 (2004) 142501.

\bibitem{Mueller2007}
P.~Mueller, I.~A. Sulai, A.~C.~C. Villari, J.~A. Alc\'antara-N\'u\~nez,
  R.~Alves-Cond\'e, K.~Bailey, G.~W.~F. Drake, M.~Dubois, C.~El\'eon,
  G.~Gaubert, R.~J. Holt, R.~V.~F. Janssens, N.~Lecesne, Z.-T. Lu, T.~P.
  O'Connor, M.-G. Saint-Laurent, J.-C. Thomas, L.-B. Wang,
  \href{https://link.aps.org/doi/10.1103/PhysRevLett.99.252501}{Phys. Rev.
  Lett.} 99 (2007) 252501.

\bibitem{Kelley1995}
J.~H. Kelley, S.~M. Austin, R.~A. Kryger, D.~J. Morrissey, N.~A. Orr, B.~M.
  Sherrill, M.~Thoennessen, J.~S. Winfield, J.~A. Winger, B.~M. Young,
  \href{https://link.aps.org/doi/10.1103/PhysRevLett.74.30}{Phys. Rev. Lett.}
  74 (1995) 30--33.

\bibitem{Tanihata1985}
I.~Tanihata, H.~Hamagaki, O.~Hashimoto, Y.~Shida, N.~Yoshikawa, K.~Sugimoto,
  O.~Yamakawa, T.~Kobayashi, N.~Takahashi,
  \href{https://link.aps.org/doi/10.1103/PhysRevLett.55.2676}{Phys. Rev. Lett.}
  55 (1985) 2676--2679.

\bibitem{F-Ne-2019}
D.~S. Ahn, N.~Fukuda, H.~Geissel, N.~Inabe, N.~Iwasa, T.~Kubo, K.~Kusaka, D.~J.
  Morrissey, D.~Murai, T.~Nakamura, M.~Ohtake, H.~Otsu, H.~Sato, B.~M.
  Sherrill, Y.~Shimizu, H.~Suzuki, H.~Takeda, O.~B. Tarasov, H.~Ueno,
  Y.~Yanagisawa, K.~Yoshida,
  \href{https://link.aps.org/doi/10.1103/PhysRevLett.123.212501}{Phys. Rev.
  Lett.} 123 (2019) 212501.

\bibitem{Thoennessen_2004}
M.~Thoennessen, \href{https://doi.org/10.1088%2F0034-4885%2F67%2F7%2Fr04}{Rep.
  Prog. Phys} 67~(7) (2004) 1187--1232.

\bibitem{GARRIDO200485}
E.~Garrido, D.~Fedorov, A.~Jensen,
  \href{http://www.sciencedirect.com/science/article/pii/S0375947403019870}{Nucl.
  Phys. A} 733~(1) (2004) 85 -- 109.

\bibitem{Casal2016}
J.~Casal, E.~Garrido, R.~de~Diego, J.~M. Arias, M.~Rodr\'{\i}guez-Gallardo,
  \href{https://link.aps.org/doi/10.1103/PhysRevC.94.054622}{Phys. Rev. C} 94
  (2016) 054622.

\bibitem{Parfenova2018}
Y.~L. Parfenova, L.~V. Grigorenko, I.~A. Egorova, N.~B. Shulgina, J.~S. Vaagen,
  M.~V. Zhukov,
  \href{https://link.aps.org/doi/10.1103/PhysRevC.98.034608}{Phys. Rev. C} 98
  (2018) 034608.

\bibitem{Tanaka2010}
K.~Tanaka, M.~Fukuda, M.~Mihara, M.~Takechi, D.~Nishimura, T.~Chinda,
  T.~Sumikama, S.~Kudo, K.~Matsuta, T.~Minamisono, T.~Suzuki, T.~Ohtsubo,
  T.~Izumikawa, S.~Momota, T.~Yamaguchi, T.~Onishi, A.~Ozawa, I.~Tanihata,
  T.~Zheng, \href{https://link.aps.org/doi/10.1103/PhysRevC.82.044309}{Phys.
  Rev. C} 82 (2010) 044309.

\bibitem{Sharov2017}
P.~G. Sharov, A.~S. Fomichev, A.~A. Bezbakh, V.~Chudoba, I.~A. Egorova, M.~S.
  Golovkov, T.~A. Golubkova, A.~V. Gorshkov, L.~V. Grigorenko, G.~Kaminski,
  A.~G. Knyazev, S.~A. Krupko, M.~Mentel, E.~Y. Nikolskii, Y.~L. Parfenova,
  P.~Pluchinski, S.~A. Rymzhanova, S.~I. Sidorchuk, R.~S. Slepnev, S.~V.
  Stepantsov, G.~M. Ter-Akopian, R.~Wolski,
  \href{https://link.aps.org/doi/10.1103/PhysRevC.96.025807}{Phys. Rev. C} 96
  (2017) 025807.

\bibitem{Charity2018}
R.~J. Charity, K.~W. Brown, J.~Oko\l{}owicz, M.~P\l{}oszajczak, J.~M. Elson,
  W.~Reviol, L.~G. Sobotka, W.~W. Buhro, Z.~Chajecki, W.~G. Lynch, J.~Manfredi,
  R.~Shane, R.~H. Showalter, M.~B. Tsang, D.~Weisshaar, J.~R. Winkelbauer,
  S.~Bedoor, A.~H. Wuosmaa,
  \href{https://link.aps.org/doi/10.1103/PhysRevC.97.054318}{Phys. Rev. C} 97
  (2018) 054318.

\bibitem{Zhukov1995}
M.~V. Zhukov, I.~J. Thompson,
  \href{https://link.aps.org/doi/10.1103/PhysRevC.52.3505}{Phys. Rev. C} 52
  (1995) 3505--3508.

\bibitem{CNO-1}
J.~G\"orres, M.~Wiescher, F.-K. Thielemann,
  \href{https://link.aps.org/doi/10.1103/PhysRevC.51.392}{Phys. Rev. C} 51
  (1995) 392--400.

\bibitem{CNO-2}
L.~V. Grigorenko, M.~V. Zhukov,
  \href{https://link.aps.org/doi/10.1103/PhysRevC.72.015803}{Phys. Rev. C} 72
  (2005) 015803.

\bibitem{Berggren1968}
T.~Berggren, Nucl. Phys. A109~(2) (1968) 265 -- 287.

\bibitem{IdBetan2002}
R.~Id~Betan, R.~J. Liotta, N.~Sandulescu, T.~Vertse,
  \href{https://link.aps.org/doi/10.1103/PhysRevLett.89.042501}{Phys. Rev.
  Lett.} 89 (2002) 042501.

\bibitem{Michel2002}
N.~Michel, W.~Nazarewicz, M.~P\l{}oszajczak, K.~Bennaceur,
  \href{https://link.aps.org/doi/10.1103/PhysRevLett.89.042502}{Phys. Rev.
  Lett.} 89 (2002) 042502.

\bibitem{Michel2003}
N.~Michel, W.~Nazarewicz, M.~P\l{}oszajczak, J.~Oko\l{}owicz,
  \href{https://link.aps.org/doi/10.1103/PhysRevC.67.054311}{Phys. Rev. C} 67
  (2003) 054311.

\bibitem{Michel2019}
N.~Michel, J.~G. Li, F.~R. Xu, W.~Zuo,
  \href{https://link.aps.org/doi/10.1103/PhysRevC.100.064303}{Phys. Rev. C} 100
  (2019) 064303.

\bibitem{Papadimitriou2011}
G.~Papadimitriou, A.~T. Kruppa, N.~Michel, W.~Nazarewicz, M.~P\l{}oszajczak,
  J.~Rotureau, \href{https://link.aps.org/doi/10.1103/PhysRevC.84.051304}{Phys.
  Rev. C} 84 (2011) 051304.

\bibitem{Michel2020}
N.~Michel, J.~G. Li, F.~R. Xu, W.~Zuo,
  \href{https://link.aps.org/doi/10.1103/PhysRevC.101.031301}{Phys. Rev. C} 101
  (2020) 031301.

\bibitem{Papadimitriou2013}
G.~Papadimitriou, J.~Rotureau, N.~Michel, M.~P\l{}oszajczak, B.~R. Barrett,
  \href{https://link.aps.org/doi/10.1103/PhysRevC.88.044318}{Phys. Rev. C} 88
  (2013) 044318.

\bibitem{Hagen2006}
G.~Hagen, M.~Hjorth-Jensen, N.~Michel,
  \href{https://link.aps.org/doi/10.1103/PhysRevC.73.064307}{Phys. Rev. C} 73
  (2006) 064307.

\bibitem{Tsukiyama2009}
K.~Tsukiyama, M.~Hjorth-Jensen, G.~Hagen,
  \href{https://link.aps.org/doi/10.1103/PhysRevC.80.051301}{Phys. Rev. C} 80
  (2009) 051301(R).

\bibitem{Sun2017}
Z.~H. Sun, Q.~Wu, Z.~H. Zhao, B.~S. Hu, S.~J. Dai, F.~R. Xu,
  \href{http://www.sciencedirect.com/science/article/pii/S0370269317302459}{Phys.
  Lett. B} 769 (2017) 227 -- 232.

\bibitem{Hu2020}
B.~Hu, Q.~Wu, J.~Li, Y.~Ma, Z.~Sun, N.~Michel, F.~Xu,
  \href{http://www.sciencedirect.com/science/article/pii/S0370269320300101}{Phys.
  Lett. B} 802 (2020) 135206.

\bibitem{zhuo2020}
Y.~Ma, F.~Xu, L.~Coraggio, B.~Hu, J.~Li, T.~Fukui, L.~D. Angelis, N.~Itaco,
  A.~Gargano,
  \href{http://www.sciencedirect.com/science/article/pii/S0370269320300617}{Phys.
  Lett. B} 802 (2020) 135257.

\bibitem{HAGEN2007169}
G.~Hagen, D.~J. Dean, M.~Hjorth-Jensen, T.~Papenbrock,
  \href{http://www.sciencedirect.com/science/article/pii/S0370269307010593}{Phys.
  Lett. B} 656~(4) (2007) 169.

\bibitem{Hagen2012}
G.~Hagen, M.~Hjorth-Jensen, G.~R. Jansen, R.~Machleidt, T.~Papenbrock,
  \href{https://link.aps.org/doi/10.1103/PhysRevLett.108.242501}{Phys. Rev.
  Lett.} 108 (2012) 242501.

\bibitem{Hagen2010}
G.~Hagen, T.~Papenbrock, M.~Hjorth-Jensen,
  \href{https://link.aps.org/doi/10.1103/PhysRevLett.104.182501}{Phys. Rev.
  Lett.} 104 (2010) 182501.

\bibitem{Hu2019}
B.~S. Hu, Q.~Wu, Z.~H. Sun, F.~R. Xu,
  \href{https://link.aps.org/doi/10.1103/PhysRevC.99.061302}{Phys. Rev. C} 99
  (2019) 061302 (R).

\bibitem{Navratil2007}
P.~Navr\'atil, V.~G. Gueorguiev, J.~P. Vary, W.~E. Ormand, A.~Nogga,
  \href{https://link.aps.org/doi/10.1103/PhysRevLett.99.042501}{Phys. Rev.
  Lett.} 99 (2007) 042501.

\bibitem{Otsuka2010}
T.~Otsuka, T.~Suzuki, J.~D. Holt, A.~Schwenk, Y.~Akaishi,
  \href{https://link.aps.org/doi/10.1103/PhysRevLett.105.032501}{Phys. Rev.
  Lett.} 105 (2010) 032501.

\bibitem{Holt2013}
J.~D. Holt, J.~Men\'endez, A.~Schwenk,
  \href{https://link.aps.org/doi/10.1103/PhysRevLett.110.022502}{Phys. Rev.
  Lett.} 110 (2013) 022502.

\bibitem{PhysRevLett.113.142501}
S.~K. Bogner, H.~Hergert, J.~D. Holt, A.~Schwenk, S.~Binder, A.~Calci,
  J.~Langhammer, R.~Roth,
  \href{https://link.aps.org/doi/10.1103/PhysRevLett.113.142501}{Phys. Rev.
  Lett.} 113 (2014) 142501.

\bibitem{PhysRevLett.113.142502}
G.~R. Jansen, J.~Engel, G.~Hagen, P.~Navratil, A.~Signoracci,
  \href{https://link.aps.org/doi/10.1103/PhysRevLett.113.142502}{Phys. Rev.
  Lett.} 113 (2014) 142502.

\bibitem{zhuo2018}
T.~Fukui, L.~De~Angelis, Y.~Z. Ma, L.~Coraggio, A.~Gargano, N.~Itaco, F.~R. Xu,
  \href{https://link.aps.org/doi/10.1103/PhysRevC.98.044305}{Phys. Rev. C} 98
  (2018) 044305.

\bibitem{zhuo2019}
Y.~Z. Ma, L.~Coraggio, L.~De~Angelis, T.~Fukui, A.~Gargano, N.~Itaco, F.~R. Xu,
  \href{https://link.aps.org/doi/10.1103/PhysRevC.100.034324}{Phys. Rev. C} 100
  (2019) 034324.

\bibitem{Coraggio2009-1}
L.~Coraggio, A.~Covello, A.~Gargano, N.~Itaco,
  \href{https://link.aps.org/doi/10.1103/PhysRevC.80.044311}{Phys. Rev. C} 80
  (2009) 044311.

\bibitem{Coraggio2009-2}
L.~Coraggio, A.~Covello, A.~Gargano, N.~Itaco, T.~Kuo,
  \href{http://www.sciencedirect.com/science/article/pii/S0146641008000410}{Prog.
  Part. Nucl. Phys.} 62~(1) (2009) 135 -- 182.

\bibitem{Roth2012}
R.~Roth, S.~Binder, K.~Vobig, A.~Calci, J.~Langhammer, P.~Navr\'atil,
  \href{https://link.aps.org/doi/10.1103/PhysRevLett.109.052501}{Phys. Rev.
  Lett.} 109 (2012) 052501.

\bibitem{Dudek1981}
J.~Dudek, Z.~Szyma\ifmmode~\acute{n}\else \'{n}\fi{}ski, T.~Werner,
  \href{https://link.aps.org/doi/10.1103/PhysRevC.23.920}{Phys. Rev. C} 23
  (1981) 920--925.

\bibitem{Bogner2003}
S.~Bogner, T.~Kuo, A.~Schwenk,
  \href{http://www.sciencedirect.com/science/article/pii/S0370157303002953}{Phys.
  Rep.} 386~(1) (2003) 1 -- 27.

\bibitem{Kuo197165}
T.~T.~S. Kuo, et~al., Nucl. Phys. A176~(1) (1971) 65 -- 88.

\bibitem{Takayanagi201161}
K.~Takayanagi,
  \href{http://www.sciencedirect.com/science/article/pii/S0375947411000145}{Nucl.
  Phys. A} 852~(1) (2011) 61 -- 81.

\bibitem{Audi2012}
G.~Audi, M.~Wang, A.~H. Wapstra, F.~G. Kondev, M.~MacCormick, X.~Xu,
  B.~Pfeiffer, \href{https://doi.org/10.1088%2F1674-1137%2F36%2F12%2F002}{Chin.
  Phys. C} 36~(12) (2012) 1287--1602.

\end{thebibliography}





\end{document}